\documentclass[english]{article}
\usepackage[T1]{fontenc}
\usepackage[latin9]{inputenc}
\usepackage{geometry}
\usepackage{amsmath}
\usepackage{amssymb}
\usepackage{setspace}
\usepackage{graphicx}
\usepackage{esint}
\makeatletter
\@ifundefined{date}{}{\date{}}
\usepackage{babel}

\makeatother

\usepackage{babel}

\begin{document}
\global\long\def\tm{\tau_{\mathrm{m}}}%
 
\global\long\def\ts{\tau_{\mathrm{s}}}%
 
\global\long\def\rout{r_{\mathrm{out}}}%
 
\global\long\def\rin{r_{\mathrm{in}}}%
 
\global\long\def\td{t_{\mathrm{d}}}%

\begin{flushleft}
\textbf{\LARGE{}Temporal support vectors for spiking neuronal networks}{\large{}}\\
\bigskip{}
\par\end{flushleft}

\begin{flushleft}
Ran Rubin\textsuperscript{1,2,3,{*}} and Haim Sompolinsky\textsuperscript{2,3,4}
\\
 \bigskip{}
\textbf{1} Center For Theoretical Neuroscience, Department of Neuroscience,
Columbia University, New-York, New-York 10032, USA \\
 \textbf{2} Edmond and Lily Safra Center for Brain Sciences, Hebrew
University, Jerusalem 91904, Israel \\
 \textbf{3} Racah Institute of Physics, Hebrew University, Jerusalem
91904, Israel \\
 \textbf{4} Center for Brain Science, Harvard University, Cambridge,
Massachusetts 02138, USA \\
 \bigskip{}
\par\end{flushleft}

\begin{flushleft}
{*} ran.rubin@mail.huji.ac.il
\par\end{flushleft}

\section*{Abstract}

When neural circuits learn to perform a task, it is often the case
that there are many sets of synaptic connections that are consistent
with the task. However, only a small number of possible solutions
are robust to noise in the input and are capable of generalizing their
performance of the task to new inputs. Finding such good solutions
is an important goal of learning systems in general and neuronal circuits
in particular. For systems operating with static inputs and outputs,
a well known approach to the problem is the large margin methods such
as Support Vector Machines (SVM). By maximizing the distance of the
data vectors from the decision surface, these solutions enjoy increased
robustness to noise and enhanced generalization abilities. Furthermore,
the use of the kernel method enables SVMs to perform classification
tasks that require nonlinear decision surfaces. However, for dynamical
systems with event based outputs, such as spiking neural networks
and other continuous time threshold crossing systems, this optimality
criterion is inapplicable due to the strong temporal correlations
in their input and output. We introduce a novel extension of the static
SVMs - The Temporal Support Vector Machine (T-SVM). The T-SVM finds
a solution that maximizes a new construct - the \emph{dynamical margin}.
We show that T-SVM and its kernel extensions generate robust synaptic
weight vectors in spiking neurons and enable their learning of tasks
that require nonlinear spatial integration of synaptic inputs. We
propose T-SVM with nonlinear kernels as a new model of the computational
role of the nonlinearities and extensive morphologies of neuronal
dendritic trees. 

\section*{Introduction}

The goal of neural learning is to adjust the weights of the inputs
to a neuron such that each of its ethologically relevant inputs will
elicit the desired output. In supervised learning, synaptic weights
are updated during an epoch in which the system has access to a set
of inputs and their desired outputs. Often however, the number of
training examples available to the system is not sufficient to fully
constrain the synaptic weights; that is, there are many weight vectors
that produce the desired outputs for the example inputs. This leads
to a fundamental problem in learning: which of these solutions is
optimal, \emph{i.e.,} offers the system with the maximal robustness
and generalization abilities, and what learning rules are likely to
converge onto the optimal solution. For classification systems, a
powerful approach has been to focus on large margin solutions, namely
solutions for which the training examples are as far as possible from
the decision surface \cite{vapnik_nature_2000}, or equivalently that
the net activation generated by the input vectors is as far as possible
from the decision threshold. In a single layer architecture \cite{rosenblatt_principles_1962,minsky_perceptrons:_1988},
the synaptic weight vector defines a decision surface in the form
of a separating hyperplane. Linear Support Vector Machines (SVM) provide
efficient algorithms for finding the hyperplane with the largest margin.
Furthermore, combining the concept of the optimal hyperplane with
the kernel method, nonlinear SVMs find an optimal solution to classification
tasks involving nonlinear decision surfaces. This latter result opens
the door for learning many real world tasks characterized by a relatively
small number of training examples and nonlinear decision surfaces
in high dimensions.

The large margin and SVM approach is appropriate for systems with
static input-output relations but is problematic for learning of spiking
neuronal networks, in which the system transforms spatio-temporal
input patterns into sequences of output spikes via threshold crossing.
This transformation can be described as a two step process. First,
inputs are integrated to produce the subthreshold membrane potential.
Second, output is generated whenever the membrane potential crosses
the neuron's threshold. At the times of threshold crossings, the membrane
potential approaches the threshold in a continuous manner, hence all
solutions necessarily have zero margin. Similar difficulties apply
to other continuous time dynamical systems that operate on temporally
continuous inputs and generate output via threshold crossing. Here
we propose a new optimality criterion for threshold crossing dynamical
systems in general and for spiking neuronal networks in particular.
This criterion is based on the novel concept of a dynamic margin,
which specifies a minimal distance from the threshold that decreases
in the vicinity of desired threshold crossing times. Using this margin,
we develop the Temporal Support Vector Machine (T-SVM) for finding
optimal synaptic weights in a single layer, spiking neural network
with linear spatial summation. We show that these solutions enjoy
greater robustness compared to solutions found by Perceptron like
learning algorithms \cite{memmesheimer_learning_2014}. Finally, we
extend T-SVM using the kernel method to construct learning algorithms
for finding optimal solutions to a class of neuronal spiking systems
with nonlinear spatio-temporal summation of inputs. We propose that
the required nonlinear kernels can be implemented by dendritic branches
with nonlinear synaptic integration, thus providing a new quantitative
framework for modeling the computational role of nonlinearities and
extensive morphologies of neuronal dendritic trees \cite{london_dendritic_2005,major_active_2013}.

\section*{Results}

\subsection*{Dynamic Margin for Spiking Neurons}

We consider the Leaky Integrate-and-Fire (LIF) model of a spiking
neuron which linearly sums $N$ input afferents and emits output spikes
when its membrane potential, $U\left(t\right)$, crosses a threshold,
$\theta$. The potential $U\left(t\right)$ is defined by 
\begin{eqnarray}
U\left(t\right) & = & \boldsymbol{\omega}^{T}\boldsymbol{x}\left(t\right)-\theta x_{\mathrm{reset}}\left(t\right)\ ,\label{eq:membrane_potential-1}
\end{eqnarray}
where $\boldsymbol{\omega}$ and $\boldsymbol{x}\left(t\right)$ are
the $N$-dimensional vectors of the synaptic efficacies and inputs
at time $t$, respectively. The input of afferent $i$ is given by
the continuous function $x_{i}\left(t\right)=\sum_{t_{i}<t}u\left(t-t_{i}\right),$
where $\left\{ t_{i}\right\} $ denote the spike train of this afferent,
and $u\left(t\right)=U_{0}\left(e^{-\frac{t}{\tm}}-e^{-\frac{t}{\ts}}\right)$
is the postsynaptic potential contributed by each spike, normalized
such that its maximal value is $1.$ The time constants $\tm$ and
$\ts$ denote the membrane and synaptic time constants, respectively.
Output spikes occur at times $t_{\mathrm{out}}$ whenever $U$ reaches
the threshold, $\theta$ . The reset term, $x_{\mathrm{reset}}\left(t\right)=\sum_{t_{\mathrm{out}}<t}u_{\mathrm{reset}}\left(t-t_{\mathrm{out}}\right)$
with $u_{\mathrm{reset}}\left(t\right)=\exp\left(-t/\tm\right)$ ensures
that $U\left(t\right)$ is reset to zero immediately after a spike. 

\begin{figure}[p]
\includegraphics[width=0.8\textwidth]{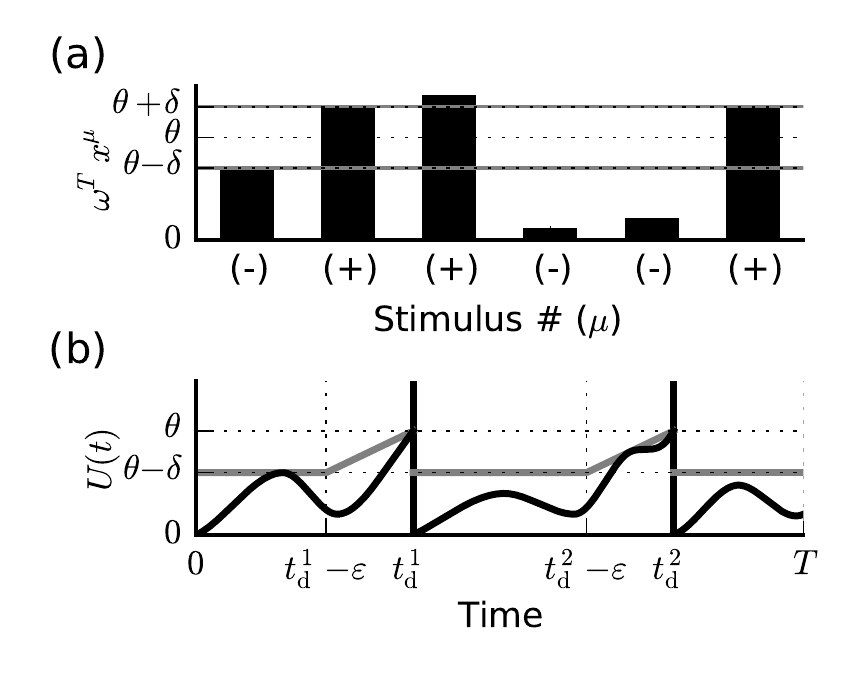}

\caption{\label{fig:1} (a){ \textbf{Margin.}
A neuron receiving static inputs. Inputs from different afferents
(elements of $\boldsymbol{x}^{\mu}$) are summed linearly, weighted
by the synaptic weights, $\omega_{i}$. The neuron classifies static
inputs by comparing the weighted sum of its inputs with a threshold
$\theta$. Patterns labeled `target' (+) are required to produce a
supra-threshold weighted sum while patterns labeled `null' (-) are
required to produce a sub-threshold weighted sum. Here all patterns
are classified correctly with a finite margin $\delta$. The optimal
solution is defined as the maximal margin solution. (b) \textbf{Dynamic
margin}. A neuron receiving dynamic inputs (elements of $\boldsymbol{x}\left(t\right)$,
eq. \ref{eq:membrane_potential-1}) and emits output spikes whenever
$U\left(t\right)$ crosses its threshold. Output spikes are required
to occur only at $\protect\td^{1}$ and $\protect\td^{2}$. The neuron
follows the required dynamics, however since the threshold must be
reached at the desired times, the margin is by definition zero. The
dynamic margin's temporal profile ($\theta-\mu\left(t\right)$, gray
line) allows $U\left(t\right)$ to be close to the threshold near
the desired crossing times while remaining far from it at other times,
thus endowing $\boldsymbol{\omega}$ with a finite dynamic margin,
$\delta$. We define the maximal dynamic margin solution as optimal.
In both (a) and (b) we assume $\left\Vert \boldsymbol{\omega}\right\Vert =1$.}}
\end{figure}

We consider a supervised learning scenario, in which the task of learning
is to modify $\boldsymbol{\omega}$ such that for a given training
input, $\boldsymbol{x}\left(t\right)$, $t\in\left[0,T\right]$, the
output spike times occur exactly at a sequence of desired output spike
times $\left\{ \td\right\} $; specifically, $U\left(t\right)$ is
required to reach $\theta$ with positive slope at each $t_{\mathrm{d}}$
while remaining strictly below $\theta$ at other times {[}Fig. \ref{fig:1}(b){]}.
As in static classification tasks there are, in general, multiple
solutions for any given task and we are confronted with the problem
of choosing the `optimal' solution among them \cite{memmesheimer_learning_2014}. 

In static classification systems, it is common to define an optimal
weight vector by requiring that, in addition to fulfilling the desired
classification of trained examples, the weight vector should obey
a large-margin optimality criterion \cite{vapnik_nature_2000}. The
margin of each weight vector is the minimal difference between the
potential generated by the input examples and the decision threshold,
where the difference is measured in units of the norm of the weight
vector, $\left\Vert \boldsymbol{\omega}\right\Vert $ {[}Fig. \ref{fig:1}(a){]}.
The optimal solution is then defined as the weight vector with the
largest possible margin. SVMs provide an efficient learning algorithm
for finding large margin linear classifiers. Furthermore, using the
Kernel method, SVMs allow extension of these algorithms to nonlinear
classification problems. 

This approach cannot be applied straightforwardly to spiking neurons.
In the model we consider, $U\left(t\right)$ must approach the threshold
$\theta$ continuously at $t_{\mathrm{d}}$. Therefore, the minimum
over time of the difference between $U(t)$ and the threshold is,
necessarily, zero {[}Fig. \ref{fig:1}(b){]}. Thus, the strong temporal
correlations in the inputs invalidate the conventional requirement
of large margins in SVMs.

Here, we offer an optimality criterion which is appropriate for neural
networks and other continuous-time threshold crossing systems. Our
criterion is based on constructing dynamic margins: For a given task
and a solution weight vector the dynamic margin, $\delta$, is defined
as 
\begin{equation}
\delta=\min_{t\ne\td}\frac{-\left(U(t)-\theta\right)}{\left\Vert \boldsymbol{\omega}\right\Vert \mu\left(t\right)}\ ,\label{eq:dynamic_margin_def}
\end{equation}
where $\mu(t)$ is a fixed temporal profile that approaches zero smoothly
with a finite slope at $t_{\mathrm{d}}$ and increases monotonically
to $1$ away from $t_{\mathrm{d}}$. Thus, qualitatively, solutions
with large dynamic margins have potentials which, away from $t_{\mathrm{d}}$,
are far away from threshold, and, close to $t_{\mathrm{d}}$, approach
the threshold with large slopes. There is a considerable freedom in
specifying the detailed shape of the temporal profile $\mu\left(t\right)$,
and different choices may be appropriate for different circumstances.
Importantly, for any given task and choice of $\mu\left(t\right)$,
every solution posses a dynamic margin $\delta>0$ {[}\emph{S. Methods}{]}.
As a concrete example, we adopt here the simple form: $\mu\left(t\right)=1$
for $t\le t_{\mathrm{d}}-\varepsilon$ and $\mu\left(t\right)=\left(t_{\mathrm{d}}-t\right)/\varepsilon$
for $t_{\mathrm{d}}-\varepsilon<t\le t_{\mathrm{d}}$ where $t_{\mathrm{d}}$
is the first desired time after time $t$ {[}See gray line in Fig.
\ref{fig:1}(b){]}. For this choice, a dynamic margin $\delta$ implies
that for all $t\notin\left[t_{\mathrm{d}}-\varepsilon,t_{\mathrm{d}}\right]$,
-$\left(U\left(t\right)-\theta\right)/\left\Vert \boldsymbol{\omega}\right\Vert \geq\delta$
and, for all $t_{\mathrm{d}}$, $\left[\mathrm{d}U\left(t_{\mathrm{d}}\right)/\mathrm{d}t\right]/\left\Vert \boldsymbol{\omega}\right\Vert \ge\delta/\varepsilon$.
The parameter $\varepsilon>0$ represents a tolerance time window
prior to $t_{\mathrm{d}}$, within which the voltage $U\left(t\right)$
approaches and reaches the threshold {[}Fig. \ref{fig:1}(b){]}.

Analogous to static large margin systems, the optimal solution, $\boldsymbol{\omega}^{\star}$
is defined, as the weight vector that possesses the maximal dynamic
margin, $\delta^{\star}$, out of all possible solutions, \textbf{\emph{$\boldsymbol{\omega}$}}.
Indeed, we show later that maximizing $\delta$ increases the robustness
of solutions to noise in the inputs, justifying the present definition
of the optimal weight vector.

\subsection*{Maximizing the Dynamic Margin}

We now develop the temporal SVM framework for finding the optimal
solution for spiking neurons based on the dynamic margin. Following
SVM theory \cite{vapnik_nature_2000} we choose a weight vector magnitude
(and threshold value) so that $\left\Vert \boldsymbol{\omega}\right\Vert \delta=1$.
Thus, maximizing the dynamic margin is equivalent to minimizing the
norm of $\boldsymbol{\omega}$ under the constraints (c.I) $U\left(t_{\mathrm{d}}\right)=\theta$
and (c.II) $\frac{\mathrm{d}U}{\mathrm{d}t}\left(t_{\mathrm{d}}\right)\ge-\frac{\mathrm{d}\mu}{\mathrm{d}t}\left(t_{\mathrm{d}}\right)$,
(c.III) $-\left(U\left(t\right)-\theta\right)\ge\mu\left(t\right)$
for all $t\ne t_{\mathrm{d}}$, 

We apply the Karush-Kuhn-Tucker theorem \cite{kuhn_nonlinear_1951}
to show that, despite the continuous nature of the inputs and constraints,
the optimal solution can be expressed as a linear sum of a small set
of select input vectors {[}\emph{S. Methods}{]},

\begin{eqnarray}
\boldsymbol{\omega}^{\star} & = & \sum_{\td}\beta_{\td}\boldsymbol{x}\left(\td\right)+\sum_{\td}\gamma_{\td}\frac{\mathrm{d}\boldsymbol{x}}{\mathrm{d}t}\left(\td\right)\label{eq:opt_w}\\
 & - & \sum_{t_{SV}}\alpha_{t_{\mathrm{SV}}}\boldsymbol{x}\left(t_{\mathrm{SV}}\right)\nonumber 
\end{eqnarray}

This set includes (i) the input vectors at some or all the desired
spike times, $\boldsymbol{x}\left(t_{\mathrm{d}}\right)$ imposing
the threshold reaching at these times. The associated Lagrange multipliers
$\beta_{d}$, can assume any real value although usually it is non-negative.
(ii) the input time derivative at some or all the desired times $\frac{\mathrm{d}\boldsymbol{x}}{\mathrm{d}t}\left(\td\right)$,
imposing the inequality on the slope of the potential at threshold;
the associated coefficients are $\gamma_{\td}\geq0$ and (iii) input
patterns, $\boldsymbol{x}\left(t_{\mathrm{SV}}\right)$, at a discrete,
problem specific, set of (non-spiking) times, $\left\{ t_{\mathrm{SV}}\right\} $,
imposing the requirement that the potential remain subthreshold at
all non-spiking times. We term this set of vectors as the Support
Vectors (SV) of the optimal solution and their associated coefficients
$\alpha_{t_{\mathrm{SV}}}>0$ as the SV coefficients. Notably, the
support vectors times, $\left\{ t_{\mathrm{SV}}\right\} $, are located
only at isolated maxima of $U\left(t\right)+\mu\left(t\right)$. 

This representation of the optimal solution allows for using well
known SVM quadratic programming techniques to evaluate the coefficients
$\alpha_{t_{\mathrm{SV}}}$, $\beta_{\td}$ and $\gamma_{\td}$ \cite{anlauf_adatron:_1989,nocedal_numerical_2006,laskov_incremental_2006,karasuyama_multiple_2009}.
We have developed an efficient bootstrap algorithm for finding these
coefficients and the set of SV times, $\left\{ t_{\mathrm{SV}}\right\} $.
This algorithm converges asymptotically to the optimal solution {[}\emph{Methods}{]}.

\subsection*{Properties of T-SVM Solutions}

\begin{figure}[p]
\includegraphics[width=0.8\textwidth]{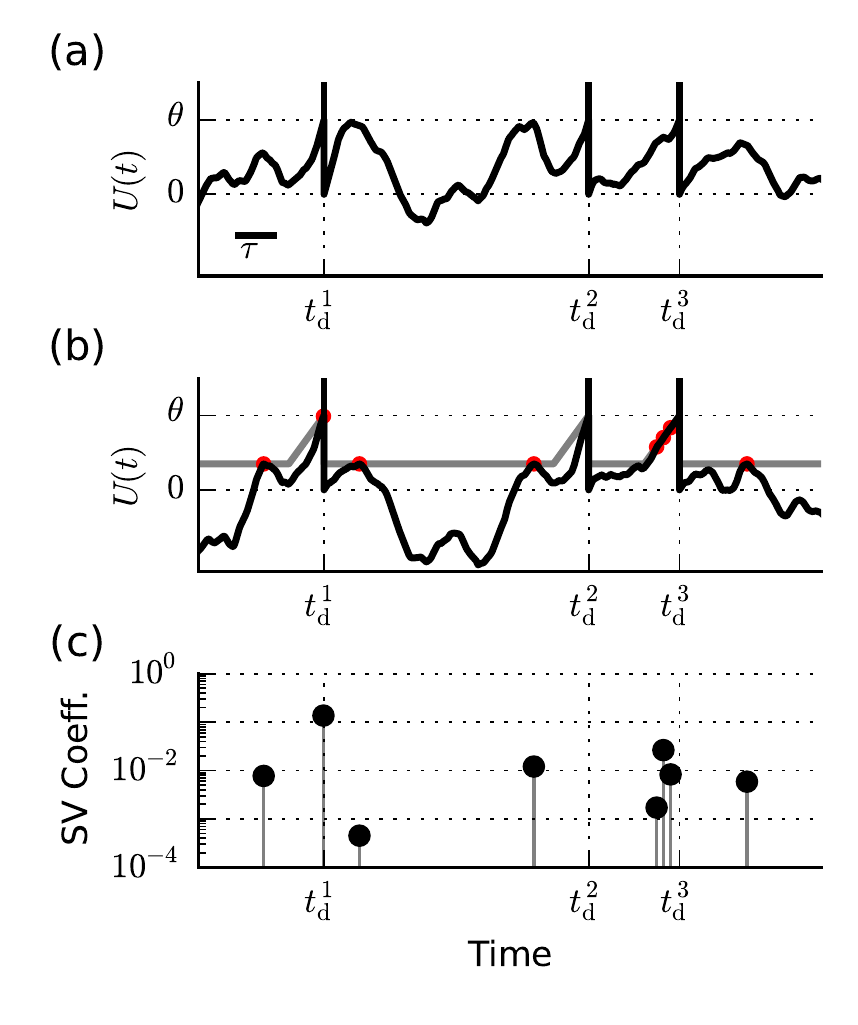}

\caption{\label{fig:traces_and_dist}{\textbf{Dynamic
margin maximization for spiking neurons.} (a)-(b) A segment of the
membrane potential traces (black) for different solutions to a single
random input-output transformation. (a) depicts a typical solution,
found by a Perceptron-like algorithm \cite{memmesheimer_learning_2014},
that possesses a close to zero dynamic margin. The characteristic
time scale of the membrane potential dynamics $\tau=\sqrt{\protect\tm\protect\ts}$
is shown. (b) depicts the optimal solution with the maximal dynamic
margin. The temporal profile of the margin, $\theta-\mu\left(t\right)$,
is depicted in gray. Red circles depict points at which $U\left(t\right)$
reaches $\theta-\mu\left(t\right)$ (c) The discrete set of support
vector coefficients for the optimal solution presented in (b). In
(b) and (c) $\varepsilon=\mbox{\ensuremath{\tau}.}$ See \emph{Methods}
for other parameters used. }}
\end{figure}

\begin{figure}[p]
\includegraphics[width=0.8\textwidth]{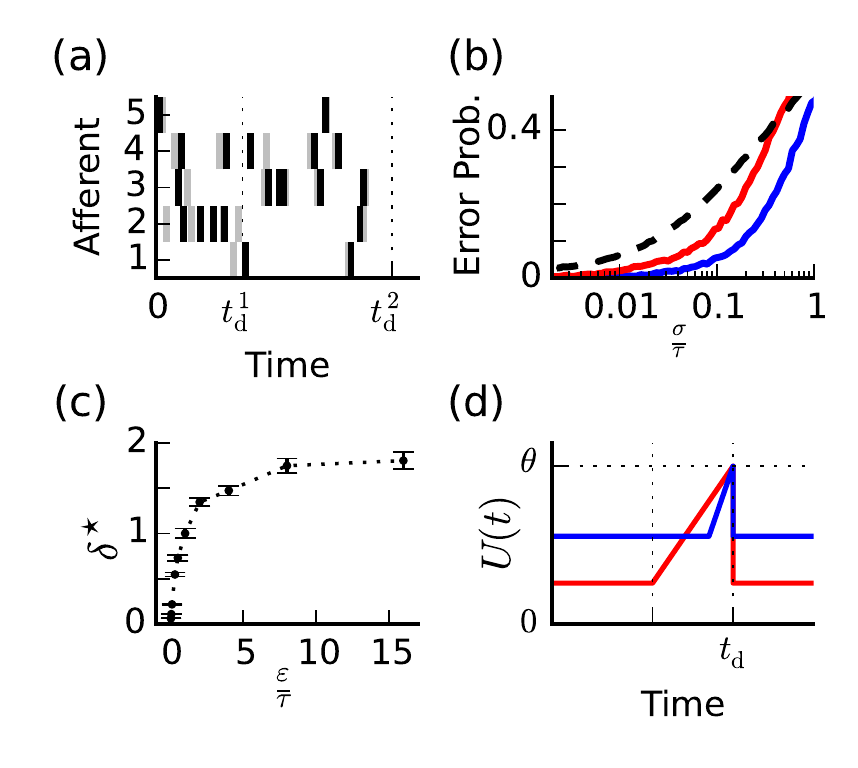}

\caption{\label{fig:jitter_and_eps}{\textbf{
Measuring the solution's robustness.} (a) Following training, the
neuron is presented with a jittered version (gray) of the learned
patterns (black). (b) Error probability per desired output spike vs.
the noise magnitude, $\sigma$, for solutions found without margin
maximization (dashed) and with margin maximization ($\varepsilon=\tau$
in red and $\varepsilon=0.3\tau$ in blue with $\tau=\sqrt{\protect\tm\protect\ts}$).
(e) Mean $\delta^{\star}$ vs. tolerance window size. Values are given
in units of the mean $\delta^{\star}$ for $\varepsilon=\tau$. Error
bars depict one standard error of the mean. (d) Temporal profile of
the dynamic margin with the mean $\delta^{\star}$ values found for
$\varepsilon=\tau$ (red) and $\varepsilon=0.3\tau$ (blue). See \emph{Methods}
for definition of error and other parameters used. }}
\end{figure}

To demonstrate the application of the T-SVM framework to spiking neurons,
we first consider the task of training the LIF neuron to spike at
desired times chosen randomly with Poisson statistics with rate $\rout$
in response to randomly chosen input spike patterns with rates $\rin$
{[}\emph{Methods}{]}. Typical solutions to this task, found without
dynamic margin maximization \cite{memmesheimer_learning_2014}, possess
very small dynamic margins {[}see example in Fig. \ref{fig:traces_and_dist}(a){]}.
In contrast, the dynamic margin of the optimal solution, $\delta^{\star}$,
is of the order of the threshold {[}Fig. \ref{fig:traces_and_dist}(b){]}.
As predicted, the SV times are located at a set of discrete time points
corresponding to maxima of $U\left(t\right)+\mu\left(t\right)$ where
$U\left(t\right)$ touches the temporal profile of the dynamic margin,
$\theta-\mu\left(t\right)$ {[}Fig. \ref{fig:traces_and_dist}(c){]}.
To demonstrate the utility of choosing the optimal solution, we examined
its robustness by presenting the neuron after learning, with noisy
versions of the trained random input-output transformation. The noise
consisted of a random Gaussian jitter with mean zero and variance
$\sigma^{2}$ added to the timing of each input spike {[}Fig. \ref{fig:jitter_and_eps}(a){]}.
The neuron's output for the optimal solution is substantially more
robust against these perturbations than the typical (non-optimal)
solution {[}Fig. \ref{fig:jitter_and_eps}(b){]}.

An important question is how the shape of $\mu\left(t\right)$ affects
the properties of the optimal solution. We explore this question by
varying the size of the margin temporal window, $\varepsilon$. We
find that the mean $\delta^{\star}$ is sensitive to the value of
$\varepsilon/\tau$, where $\tau=\sqrt{\tm\ts}$ is the characteristic
correlation time of the inputs; $\delta^{\star}$ approaches zero
as $\varepsilon/\tau\rightarrow0$ and saturates to a finite value
for $\varepsilon/\tau\gg1$ {[}Fig. \ref{fig:jitter_and_eps}(c){]}.
However, despite the increase in dynamic margin for $\varepsilon/\tau\gtrsim1$,
we find that the performance in the presence of noise reaches an optimum
at intermediate values, $\varepsilon/\tau\approx0.3$ {[}Fig. \ref{fig:jitter_and_eps}(b),
blue line{]}. This optimum reflects the opposing effects of increasing
$\varepsilon$. While increasing $\varepsilon$ increases the margin
at times away from the desired spikes, it decreases the minimum allowed
slope of the voltage as it crosses threshold at the desired times
{[}Fig. \ref{fig:jitter_and_eps}(d){]}. Thus the optimal value of
$\varepsilon/\tau$ represents a tradeoff between the robustness of
desired spikes (reflected by the slope at $t_{\mathrm{d}}$) and the
distance of the membrane potential from the decision threshold away
from $t_{\mathrm{d}}$.

\begin{figure}[p]
\includegraphics[width=0.8\textwidth]{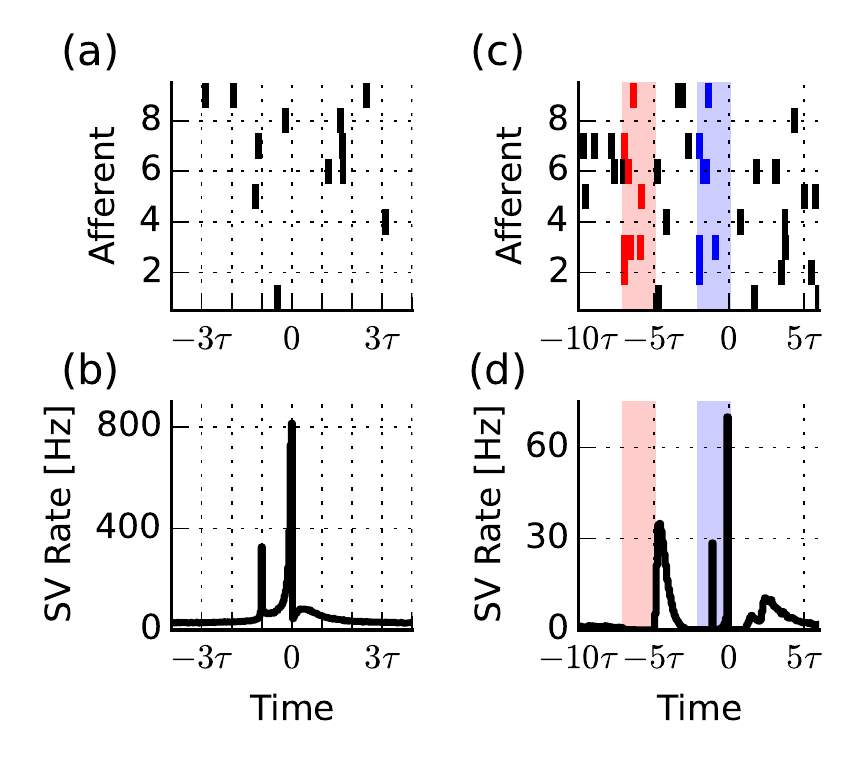}

\caption{\label{fig:sv_dist}{\textbf{Distribution
of SV times.} (a) An example of a random input spike train relative
to a desired time $t_{\mathrm{d}}$ ($t=0$). Spike times are drawn
from independent Poisson processes. (b) Distribution of support vector
times relative to $t_{\mathrm{d}}$ for random transformations. (c)
Structured input patterns. The neuron is presented with a stream of
random Poisson spikes (black) embedded with pairs of repeating stimuli
composed of two similar patterns (red and blue). The neuron's task
is to fire only at the end of the blue pattern ($t=0$). (d) Distribution
of support vector times relative to the desired time ($t=0$) for
the structured input patterns presented in (c). See \emph{Methods}
for other parameters used.}}
\end{figure}

It is interesting to examine the distribution of $\left\{ t_{\mathrm{SV}}\right\} $
relative to $t_{\mathrm{d}}$. For random patterns this distribution
mostly reflects the shape of $\mu\left(t\right)$ and the temporal
correlation function of the input potentials and post synaptic reset
{[}Fig. \ref{fig:sv_dist}(a)-(b){]}. A high density of SVs just prior
to $t_{\mathrm{d}}$ ensures the high slope of $U\left(t\right)$
prior to and at $t_{\mathrm{d}}$. The secondary peak is located exactly
at $t_{\mathrm{d}}-\varepsilon$ where the time derivative of $\mu\left(t\right)$
is discontinuous. For tasks with non-random spike patterns, the distribution
of SV times is also affected by the spatio-temporal structure of the
input spikes and desired output spikes. An example is shown in Figure
Fig\emph{.} \ref{fig:sv_dist}. Inputs consist of two similar but
not identical stimuli, a `null' pattern and a `target' pattern {[}\emph{Methods,
}Fig\emph{.} \ref{fig:sv_dist}(c), red and blue respectively{]}.
The neuron's task is to remain silent in response to the `null' pattern
and fire at the end of the `target' pattern. These patterns are separated
by a constant time interval. They repeat at random times and are embedded
by a random background Poisson input spikes, making the task of responding
to one but not the other input nontrivial. The learning rule finds
a large margin solution to this task. The distribution of SV times
exhibits a large concentration of SVs immediately after the end of
the `null' pattern. These SVs prevent erroneous output spikes in response
to this pattern.

\begin{figure}[p]
\includegraphics[width=1\textwidth]{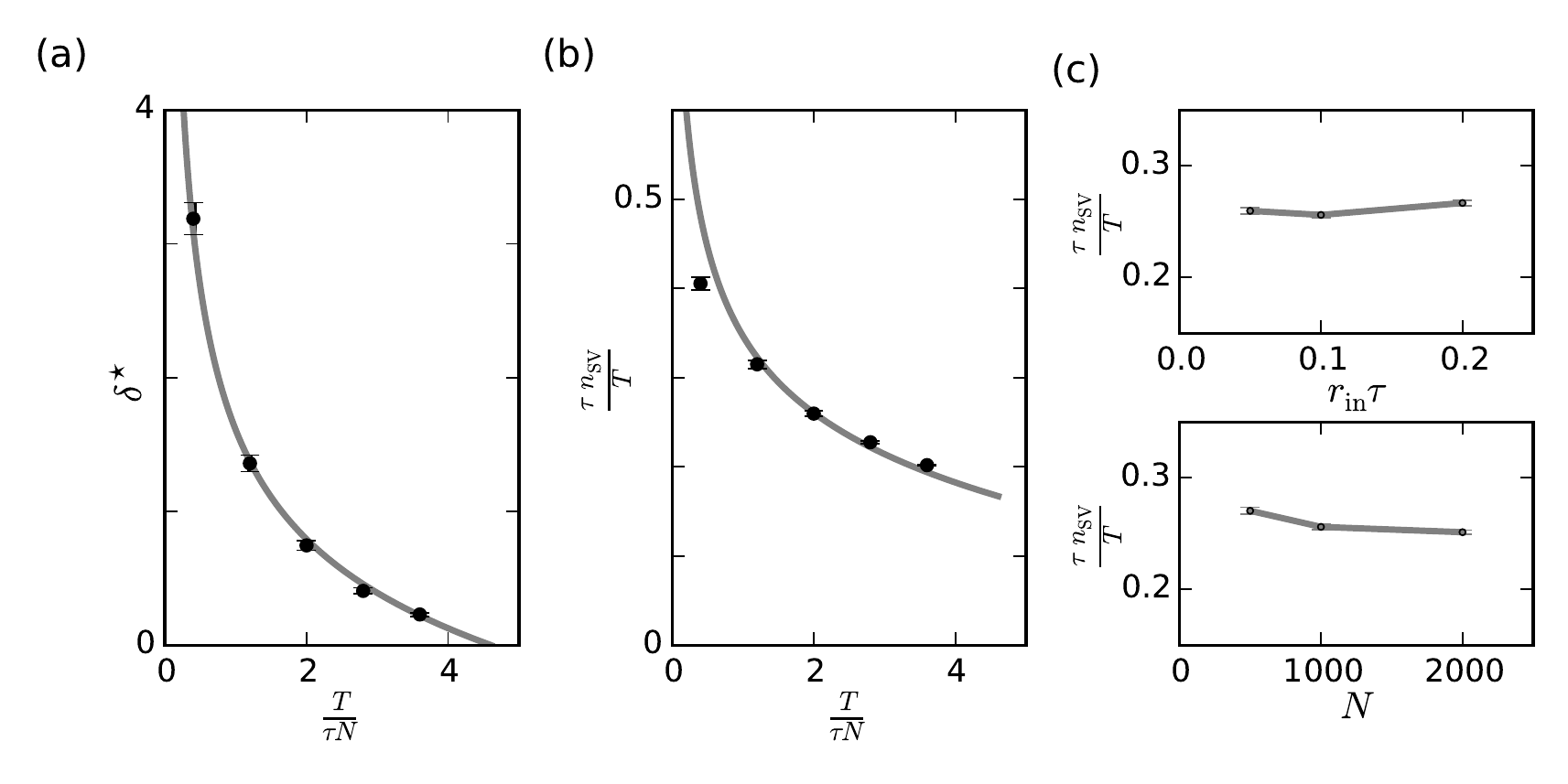}

\caption{\label{fig_LIF_theory}(a) Mean maximal dynamic margin, $\delta^{\star}$,
vs. learned pattern duration, $T$. $\delta^{\star}$ is shown in
units of the standard deviation of $x_{i}\left(t\right)$. (b) Mean
number of SVs per unit time vs. learned pattern duration, $T$. In
(a) and (b) black error bars depict simulation data while gray lines
depict the approximated theoretical predictions. (c) Mean number of
support vectors per unit time vs. mean number of input spikes in time
$\tau$ for constant $N=1000$ (top) and vs. $N$ for constant $\protect\rin\tau=0.1$
(bottom). In all panels $\varepsilon=\mbox{\ensuremath{\tau}, \ensuremath{\tau=0.01\sec}},$
$\protect\rout\tau=0.05,\ \protect\tm/\protect\ts=4$. In (a) and
(b) $N=1000,$ $\protect\rin\tau=0.05$. In (c) $T/\tau N=2$.}
\end{figure}

Finally we study how the properties of the maximal dynamic margin
solution change with the parameters of the learning task. We find
that $\delta^{\star}$ decreases with increasing duration, $T,$ of
the learned input-output sequence {[}Fig. \ref{fig_LIF_theory}(a){]}.
This is expected since as $T$ increases the weight vector must satisfy
an increasing number of constraints. In fact, $\delta^{\star}$ approaches
zero at a critical value of $T$, denoting the maximal capacity, beyond
which the desired input-output transformation cannot be implemented
by an LIF neuron. In addition to the capacity, we are able to analytically
calculate {[}\emph{S. Methods}{]} the maximal dynamic margin and the
number of support vectors per unit time {[}Fig. \ref{fig_LIF_theory}(a)-(b){]}.
Importantly, the number of SVs per unit time does not increase with
the number of input spikes as $\rin$ or $N$ increase {[}Fig. \ref{fig_LIF_theory}(c){]}.
This demonstrates that a finite number of SVs per synapse is sufficient
even when time is continuous and the total rate of input spikes tends
to infinity. For example, for intermidated loads and $\rout\tau=0.05,$
the mean number of support vectors is approximatly 0.25 SV's per time
$\tau$. For a neuron with, $\tau\sim10\mathrm{msec}$, and $\rout\sim5\mathrm{Hz}$
this implys about 25 support vectors per 1 sec. of learned input-output
transformation independent of $N$, $T$ and $\rin$.

\subsection*{Nonlinear Computation Using the Kernel Method}

Many interesting tasks require nonlinear summation of inputs. Our
approach can be extended to include such nonlinearities. To do so,
we first note that when the optimal weights, Eq. \ref{eq:opt_w},
are substituted, the linear potential can be written as 

\begin{equation}
\boldsymbol{\omega}^{T}\boldsymbol{x}\left(t\right)=\sum_{\ell}a^{\ell}\mathcal{K}\left(\boldsymbol{x}^{\ell},\boldsymbol{x}\left(t\right)\right)\ ,\label{eq:Kernel_linear}
\end{equation}
where, $\mathcal{K}\left(\boldsymbol{x}^{\ell},\boldsymbol{x}\left(t\right)\right)=\boldsymbol{x}^{\ell T}\boldsymbol{x}\left(t\right)$
and for the optimal weights, $\left\{ \boldsymbol{x}^{\ell}\right\} $
is the set of training vectors $\left\{ \boldsymbol{x}\left(\td\right),\frac{\mathrm{d}\boldsymbol{x}}{\mathrm{d}t}\left(\td\right),\boldsymbol{x}\left(t_{\mathrm{SV}}\right)\right\} $
and the coefficients $a^{\ell}$ are the corresponding Lagrange coefficients
(see \eqref{eq:opt_w}). Extending the linear potential to the non-linear
case, we assume that the membrane potential has the from

\begin{equation}
U\left(t\right)=\sum_{l}a^{\ell}\mathcal{K}\left(\boldsymbol{x}^{\ell},\boldsymbol{x}\left(t\right)\right)-\theta x_{\mathrm{reset}}\left(t\right)\ ,\label{eq:Kernel}
\end{equation}
where $a^{\ell}$ and the $N$-dimensional vectors, $\boldsymbol{x}^{\ell}$,
are adjustable parameters and $\mathcal{K}\left(\boldsymbol{v},\boldsymbol{u}\right)$
is a nonlinear symmetric positive definite kernel \cite{vapnik_nature_2000}.
An example is a polynomial kernel $\mathcal{K}\left(\boldsymbol{v},\boldsymbol{u}\right)=(\boldsymbol{v}^{T}\boldsymbol{u})^{d}$
generalizing the linear case $d=1$ . As before, for a training input
$\boldsymbol{x}\left(t\right)$, and desired threshold crossing times,
$\left\{ \td\right\} $, $U\left(t\right)$ is required to reach threshold
at and only at $\td$. For a fixed kernel form, the goal of learning
is to determine the appropriate `template inputs', $\boldsymbol{x}^{\ell}$,
and coefficients $a^{\ell}$. Similar to the kernel extension of SVM,
the T-SVM optimization can be extended to the nonlinear case {[}\emph{S.
Methods}{]}. The resulting optimal solutions are such that the $a^{\ell}$
coefficients correspond to the Lagrange multiplier coefficients and
optimal `template inputs', $\boldsymbol{x}^{\ell}$, correspond to
$\left\{ \boldsymbol{x}\left(\td\right),\boldsymbol{x}\left(t_{\mathrm{SV}}\right)\right\} $
where $t_{\mathrm{SV}}$ are a set of non-spiking times (In the nonlinear
case we do not explicitly use the time derivatives of the input vectors,
see \emph{S. Methods} for details). 

\begin{figure}[p]
\includegraphics[width=0.8\textwidth]{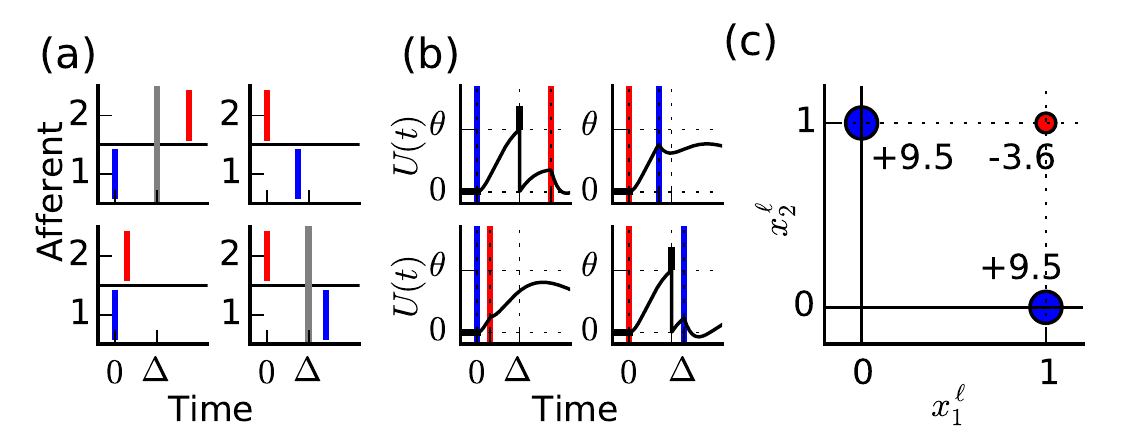}

\caption{\label{fig:Temporal-XOR}\textbf{Non Linear Temporal Computation}
(a) The neuron receives a single spike from each of its two afferents
at times $t_{1}$ and $t_{2}$. The neuron is required to fire at
a time $\Delta$ after the first input ($t=0)$ if $\left|t_{1}-t_{2}\right|>\Delta$
(top left and bottom right, gray line) and remain silent if $\left|t_{1}-t_{2}\right|<\Delta$
(bottom left and top right). A neuron with a quadratic kernel,~$\mathcal{K}\left(\boldsymbol{v},\boldsymbol{u}\right)=\left(\boldsymbol{v}^{T}\boldsymbol{u}\right)^{2}$,
is trained with 9 patterns with $t_{1}-t_{2}$ in the range of $\pm1.25\Delta$.
(b) Response of trained neuron to `target' patterns (top left and
bottom right) and `null' patterns (bottom left and top right). Blue
and red lines depict the input spikes times $t_{1}$ and $t_{2}$
respectively. As required the neuron emits an output spike at time
$\Delta$ in the target patterns and remains silent in the null patterns.
(c) Circles depict the resulting `template inputs', $\boldsymbol{x}^{\ell}$,
and their corresponding coefficients for the optimal solution. The
size of the circles is proportional to the coefficient's value. The
resulting subthreshold potential is given by $U_{\mathrm{sub}}(t)=9.5\left(x_{1}(t)^{2}+x_{2}(t)^{2}\right)-3.6\left(x_{1}(t)+x_{2}(t)\right)^{2}$.}
\end{figure}

We apply the nonlinear T-SVM to a temporal variant of the XOR problem,
a hallmark of nonlinear computation {[}Fig. \ref{fig:Temporal-XOR}{]}.
A spiking neuron receives inputs from afferents that fire a single
spike during a trial. The neuron has to fire at a time $\Delta$ after
the firing of any of its inputs, provided that no additional input
arrived during the delay period {[}Fig. \ref{fig:Temporal-XOR}(a){]}.
Thus, the salient feature that the neuron must be tuned to is an inter-spike-interval
larger than $\text{\ensuremath{\Delta}}$ in the sequence of incoming
spikes. This task cannot be performed by a linear summation of inputs
since each input spike, in isolation, must be able to elicit an output
spike, whereas the proximal firing of two afferents must leave the
neuron subthreshold. Using T-SVM learning we show that a neuron with
two input afferents and a quadratic kernel, $\mathcal{K}\left(\boldsymbol{v},\boldsymbol{u}\right)=(\boldsymbol{v}^{T}\boldsymbol{u})^{2}$,
can implement the task {[}Fig. \ref{fig:Temporal-XOR}(c)-(d){]}.
In this case, the optimal solution can be expressed by 3 simple template
input vectors, $\boldsymbol{x}^{1}=\left(1,0\right)$, $\boldsymbol{x}^{2}=\left(0,1\right)$
and $\boldsymbol{x}^{3}=\left(1,1\right)$. The coefficients for $x^{1}$
and $\boldsymbol{x}^{2}$ are positive and guarantee that each input
spike in isolation drives the neuron to fire {[}Fig. \ref{fig:Temporal-XOR}(b),
(c){]}. The negative SV coefficients associated with $\boldsymbol{x}^{3}$
guarantee that an additional incoming spike prior to $\Delta$ interacts
nonlinearly with the first spike, curtailing the threshold crossing
of $U\left(t\right)$ {[}Fig. \ref{fig:Temporal-XOR}(b), (d){]}.

\subsection*{A New Framework for Dendritic Computation}

\begin{figure*}[p]
\includegraphics[width=0.8\textwidth]{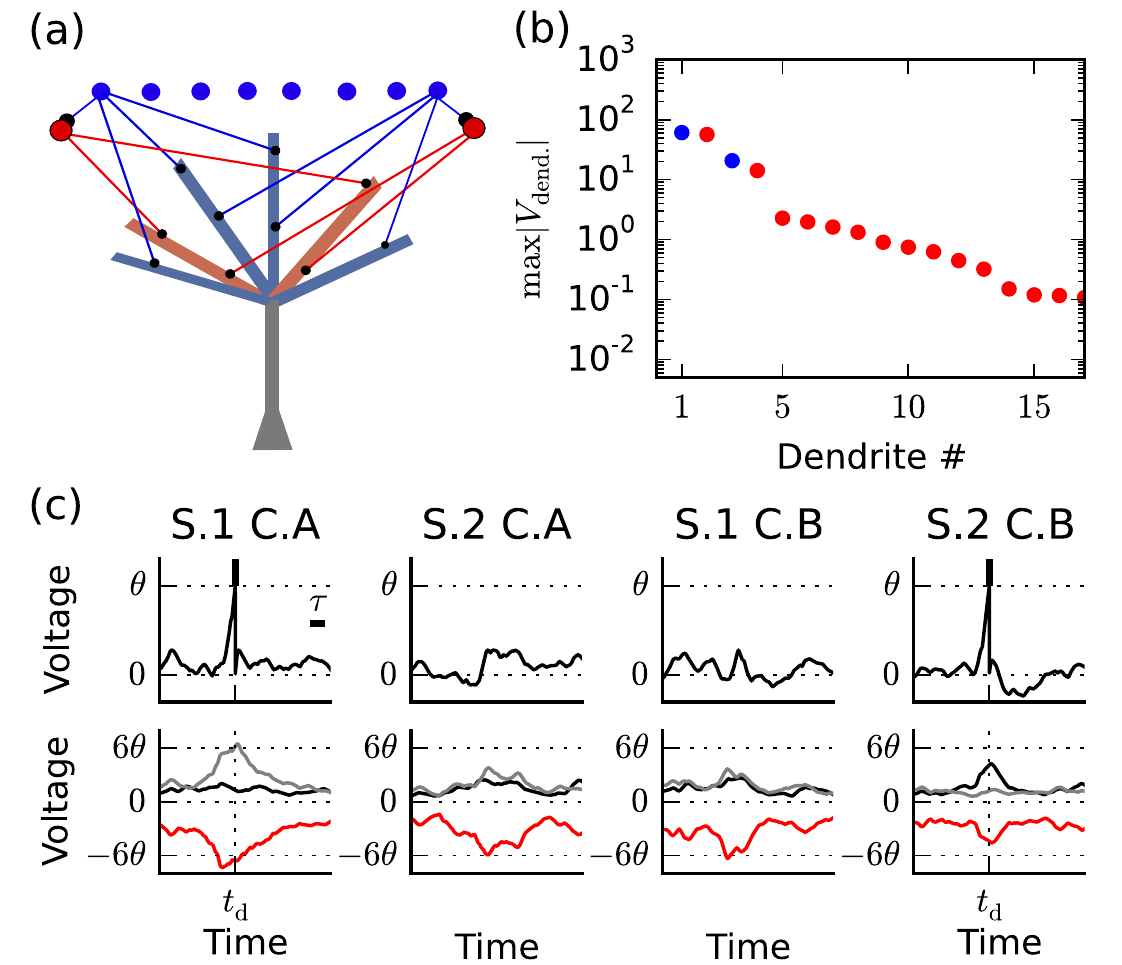}

\caption{\label{fig:Neuronal-architecture-for-nonlinear-kernels}\textbf{Neuronal
architecture for implementing temporal nonlinear kernels.} (a) Input
afferents (blue circles) project both excitatory (blue) and inhibitory
(red), connections onto the neuron's dendritic tree. Inhibitory inputs
from each input afferent may be implemented via feed forward inhibition
(red circles). Each input afferent potentially innervates multiple
dendritic branches. Inputs are continuously summed nonlinearly within
each branch implementing the nonlinear kernel ($\mathcal{K}$ in eq.
\ref{eq:Kernel}). The contribution from each branch is linearly summed
in the neuron's soma. (b) Maximal contribution of each dendrite to
the membrane potential of a neuron with non linear dendrites implementing
T-SVM with a quadratic kernel trained on a context dependent respone
task (see text). For a quadtaic kernel each dendrite receives exclusively
excitatory (blue) or inhibitory (red) input. Dendrites are sorted
according to their maximal contribution and the first 17 dendritic
branches contribute 99\% of the total membrane potential. (c) Top:
Voltage traces of the trained neuron. As required, the neuron responds
to stimulus $S.1$ with one spike at time $\protect\td$ and remain
silent in response to stimulus $S.2$, while in context $C.A$ and
interchanges its responses while in context $C.B$. Bottom: Traces
depict the summed voltage contribution of the two first dendrites
pairs (see (b)) in black and gray respectively, and the summed contribution
of the remaining dendrites (red). The output spike is driven by dendrites
1 and 2 for pattern $S.1\ C.A$, and by dendrites 3 and 4 for pattern
$S.2\ C.B$. The pair (1,2) implements a non-linear AND operation
for patterns $S.1$ and $C.A$ while the pair (3,4) implements an
AND operation for the patterns $S.2$ and $C.B$. The total contribution
of the other dendrites is largely non-selctive inhibition; however
it is important for shaping the precisely timed response of the neuron.
See \emph{Methods }for parameters used.}
\end{figure*}

The nonlinearities of synaptic integration in the dendritic trees
of many neuron types are well established \cite{london_dendritic_2005,major_active_2013}.
However, their role in neural information processing is still an
open question. According to one proposal \cite{poirazi_impact_2001,polsky_computational_2004},
the dendritic morphology endow pyramidal cells with a functional `two-layer'
Perceptron architecture, in which each nonlinear dendritic branch
acts like a sigmoidal unit in the hidden layer of the multi-layer
Perceptron. Here we propose that neurons may be able to utilize dendritic
nonlinearities to perform nonlinear computation in the spatio-temporal
domain by emulating T-SVM with nonlinear kernels. Specifically, the
nonlinear synaptic potential, \eqref{eq:Kernel}, can be interpreted
as a weighted sum of contributions from individual dendritic branches.
For a kernel of the form $\mathcal{K}\left(\boldsymbol{v},\boldsymbol{u}\right)=f\left(\boldsymbol{v}^{T}\boldsymbol{u}\right),$
the function $f$ may represent the dendritic nonlinearity and the
components of the vectors $\boldsymbol{x}\left(t_{\mathrm{d}}\right)$
, and $\boldsymbol{x}\left(t_{\mathrm{SV}}\right)$ can be interpreted
as the efficacies of the synapses that innervate the corresponding
dendrite. The inputs from different dendrites are then summed linearly
at the soma with the SV coefficients representing the coupling strength
between a dendrite and the soma {[}Fig. \ref{fig:Neuronal-architecture-for-nonlinear-kernels}(a){]}.
Interestingly, emulating T-SVM with the simple quadratic kernel implies
that the contribution from each dendritic branch is either excitatory
or inhibitory. Higher order kernels allow for mixing of excitatory
and inhibitory inputs on the same dendrite.

To demonstrate the dendritic interpretation of kernel T-SVM, we consider
a task in which the same sensory inputs should elicit different responses
based on some contextual information. Consider, for example, a neuron
that receives inputs from two groups of neurons representing stimulus
and context respectively. In contextual state $C.A$ the neuron is
required to respond to stimuli $S.1$ and $S.2$ with given sets of
desired output spike times respectively. In context $C.B$ the desired
responses are interchanged. Thus, for the desired output times the
neuron must perform an XOR like operation which cannot be implemented
in a linear architecture. To understand how this computation can be
accomplished with a dendritic tree structure we analyze the simple
case where the context dependent response task consists of a single
desired output spike {[}Fig. \ref{fig:Neuronal-architecture-for-nonlinear-kernels}(b)-(c){]}.
Using a quadratic kernel, $\mathcal{K}\left(\boldsymbol{v},\boldsymbol{u}\right)=\left(\boldsymbol{v}^{T}\boldsymbol{u}\right)^{2}$,
the T-SVM algorithm can find a robust solution for this task {[}Fig.
\ref{fig:Neuronal-architecture-for-nonlinear-kernels}(c), top{]}.
In this example 25 stimulus afferents and 25 context afferents impinge
on a postsynaptic neuron. 17 dendritic branches analog to 17 Lagrange
coefficients and their associated input templates, are sufficient
to approximate the optimal solution with 99\% accuracy {[}Fig. \ref{fig:Neuronal-architecture-for-nonlinear-kernels}(b){]},
contribution from other Lagrange coefficients is negligable. The nonlinear
computation relevant for each output spike is distributed onto different
sets of dendrites. Specifically, two pairs of dendrites yields the
most significant contributions to the total membrane potential {[}Fig.
\ref{fig:Neuronal-architecture-for-nonlinear-kernels}(b){]}. However
the total activity of one pair (dendrites 1-2) is modulated only in
response to pattern $S.1\ C.A$ while the total activity of the other
pair (dendrites 3-4) is modulated in response to pattern $S.2\ C.B$
{[}Fig. \ref{fig:Neuronal-architecture-for-nonlinear-kernels}(c),
bottom{]}. Thus each pair implements a logical AND operation and their
output is summed linearly at the soma. It is interesting to note that
the scale of activation of individual dendrites is much greater then
the scale of the membrane potential itself (compare the vertical scales
of top and bottom in Fig. \ref{fig:Neuronal-architecture-for-nonlinear-kernels}(c)).
Thus, similar to balanced networks \cite{van_vreeswijk_chaos_1996,vreeswijk_chaotic_1998},
the neuron's net membrane potential results from the fluctuations
in the large excitatory and inhibitory components that largely cancel
each other. When more then one desired output spike is learned, the
solution structure is more complex: we find some excitatory dendrites
without a corresponding inhibitory dendrite, and that desired spikes
may be implemented by more then one dendrite pair.

\section*{Discussion}

Supervised learning in spiking networks has been the subject of several
recent studies (for a review see \cite{gutig_spike_2014}). In a previous
work \cite{memmesheimer_learning_2014} we have developed a Perceptron
like synaptic learning algorithm for a neuron that is trained to generate
a sequence of spikes in given times. While the algorithm converges
to a solution when there is one, it does not in general finds a particularly
robust solution or one with good generalization abilities. Furthermore,
this learning rule does not generalize to tasks requiring nonlinear
summation of synaptic inputs. 

Here we develop learning rules for finding the optimal weight vectors
for spiking neurons extending the theory of large margin systems.
Large margin systems and in particular Support Vector Machines have
been attractive model systems in Machine and Statistical Learning,
due to the robustness of their output, the good generalization capabilities,
and the existence of efficient learning algorithms for both the linear
and nonlinear systems. This motivates the question whether spiking
neuronal circuits can realize similar functionalities. A basic difficulty
is the fact that the potential of a spiking neuron approaches the
spiking threshold smoothly, implying that the difference between the
potential and the decision boundary (the analog of margin in static
systems) approaches zero as the neuron fires. To solve this problem
we have introduced the new concept of dynamic margin, Importantly,
the dynamic margin takes into account the temporal correlations of
the system's dynamics. The application of dynamic margin maximization
and the T-SVM framework is not limited to spiking neuron models and
can be generalized to any dynamical system in which output is based
on threshold crossing {[}\emph{S. Methods}{]}. It allows for robust
threshold crossings at desired times while maintaining a large gap
between the system's state variable (membrane potential in the case
of neurons) and the decision threshold at other times. Despite the
continuous time dynamics, the synaptic weight vector maximizing the
dynamic margin can be described using a finite set of parameters,
associated with the inputs at the desired output spike times and select
times during quiescence; these parameters are analogous to the support
vector representation of large margin solutions in static Machine
Learning systems. The number of required parameters is proportional
to the total duration of the required temporal output spike trains.
Indeed, we have shown that the solution with the maximal dynamic margin
enjoys greater robustness to input noise. In this work, we have adopted
a simple piecewise linear form of the temporal shape of the profile
$\mu(t)$ underlying the definition of the dynamic margin, \ref{eq:dynamic_margin_def}.
It would be interesting to explore the properties of other temporal
profiles and determine how the margin, the robustness to noise, and
generalization abilities are affected by the choice of $\mu(t)$.
Second, it would be interesting to extend the theory and algorithms
to the more biologically relevant case, where the desired times are
specified only up to some preassigned precision \cite{memmesheimer_learning_2014}.
Finally, the theory should be extended to allow finding minimal error
solutions for cases where the desired input-output transformation
is not implementable.

Importantly, we extended the learning capabilities of our spiking
neuron model to nonlinear computations by generalizing the kernel
method of Support Vector Machines to dynamical threshold crossing
systems. We have demonstrated that kernels with relatively simple
quadratic nonlinearities are sufficient to perform nonlinear tasks
such as temporal and spatial XOR computations and context dependent
responses. Indeed quadratic nonlinearities have been shown to account
for the nonlinear operation of motion sensing neurons and Complex
Cells in visual cortex \cite{hassenstein1956systemtheoretische,adelson1985spatiotemporal,emerson1992directionally}. 

In the present work, we have assumed that learning works by modifying
the synaptic (linear and nonlinear) efficacies, whereas the temporal
integration of synaptic inputs was assumed fixed. However, the T-SVM
framework can be extended to more general adaptive dynamical systems
in which the temporal filtering of the inputs is adjustable and can
be optimized by learning {[}see example in\emph{ S. Methods}{]}. Using
the kernel method, this extension can be generalized to neurons with
nonlinear spatio-temporal kernels allowing for nonlinear interactions
between different spatial components of the input and inputs at different
times. We leave the investigation of these models and their biological
interpretation for future work. 

We have proposed that the underlying kernel nonlinearities are implemented
in biological circuits by the nonlinear synaptic integration of dendritic
branches. Our proposed function for dendritic nonlinearities share
similarity with the pioneering work of Mel \emph{et al. \cite{poirazi_impact_2001,polsky_computational_2004,jadi_augmented_2014}}.
The difference between the two proposed functional architectures is
similar to the difference between a two layer perceptron and that
of a Kernel SVMs. The advantage of the latter is the existence of
an efficient learning algorithm that is guaranteed to converge to
the (optimal) solution provided that such a solution exists. In the
present context, our proposal differs from earlier models in that
it takes into account explicitly the dynamic nature of the decision
making of spiking neuron, whereas the multilayer perceptron model
is a rate based, largely static input-output system. Furthermore,
using the SV representation, our proposal provides insight into the
contributions of the different dendrites to the performance of the
task, by encoding in their synapses specific template input vectors
representing the stream of training inputs during spiking and nonspiking
times. Such interpretation is lacking for solutions derived by generic
gradient based approaches to learning in a two-layer perceptron and
the dendritic architectures that mimic them. Nevertheless, the two-layer
perceptron model is advantageous to our model in certain aspects.
In particular, in our SVM model, to allow for successful learning,
each presynaptic source should be in general allowed to innervate
all branches; whereas the input architecture can be restricted in
learning the two-layer perceptron model. Further studies of learning
in neuron models with realistic morphologies learning to perform interesting
computations are needed in order to shed further light on the important
question of the functional role of the extensive nonlinearities and
morphologies of neurons' dendritic trees. Finally, the learning algorithms
presented here are designed to find and reveal the properties of optimal
weight vectors of trained spiking neurons. At present, SVM like learning
algorithms, as the one presented here are not a plausible model of
the biological process of learning. Implementing large margin solutions
of these algorithms by a neurally plausible online learning algorithm
remains an important open challenge for computational neuroscience. 

\section*{Methods}

\textbf{T-SVM Algorithm:} Here we describe the algorithm used to find
the maximal dynamic margin solution. The algorithm consists of two
iterative stages: Sampling stage and Optimization Stage. In the sampling
stage we construct a discrete set of times (and corresponding input
patterns) on which optimization should be performed. In the optimization
stage we use quadratic programing algorithms to find the optimal value
of $\theta$ and the Lagrange multipliers.

The algorithm is as follows: 
\begin{enumerate}
\item \textbf{Initialization}: Choose a set of sampled times $\left\{ t_{\mathrm{SV}}\right\} $
arbitrarily. 
\item \textbf{Optimization}: Maximize the dynamic margin using only the
current set $\left\{ t_{\mathrm{SV}}\right\} $ by using a quadratic
programing algorithm. 
\item \textbf{Sampling}

\begin{enumerate}
\item Present all the input patterns to the neuron using the neuron's dynamics
and the current set of parameters. 
\item Construct a set of new sampled times, $\left\{ t_{\mathrm{new}\text{ }}\right\} $
consisting of the times of the maximal value of $U$$\left(t\right)+\mu\left(t\right)$
in every time segment in which $U\left(t\right)>\theta-\mu\left(t\right)$\textbf{.} 
\item For every $t_{\mathrm{new}}\notin\left\{ t_{\mathrm{SV}}\right\} $
add $t_{\mathrm{new}}$ to the set $\left\{ t_{\mathrm{SV}}\right\} $.
$t_{\mathrm{new}}$ is considered to be in $\left\{ t_{\mathrm{SV}}\right\} $
if $\exists t_{\mathrm{SV}}:\ \left|t_{\mathrm{SV}}-t_{\mathrm{new}}\right|<\epsilon_{t}$
where $\epsilon_{t}$ is an arbitrarily small time constant. 
\end{enumerate}
\item \textbf{Stopping Criteria: }If no new times were added to $\left\{ t_{\mathrm{SV}}\right\} $,
stop. 
\item \textbf{Return to step 2.} 
\end{enumerate}
For any $\epsilon_{t}>0$ this algorithm samples at most $T/\epsilon_{t}$
points in time. Thus for any $\epsilon_{t}>0$ the algorithm will
converge in a finite number of iterations. For $\epsilon_{t}\rightarrow0$
this algorithm will asymptotically converge to the optimal solution.

\textbf{Desired outputs for random patterns:} To avoid pathologies,
output spikes are not allowed between $t=0$ and $t=\tm$. For $t\in\left(\tm,T\right)$
spikes are drawn from a Poisson process with rate $\rout/\left(1-\frac{\tm}{T}\right)$.\textbf{ }

\textbf{Definition of error due to jitter: }We observed that the mean
inaccuracy of desired spikes in error free time segments is small
compared to $\tau$ and approximately proportional to the standard
deviation of the jitter (not shown). Thus, we define the neuron's
output in the $n$'th desired spike time, $t_{\mathrm{d}}^{n}$, as
erroneous if the number of output spikes in $\left[\left(t_{\mathrm{d}}^{n-1}+t_{\mathrm{d}}^{n}\right)/2,\ \left(t_{\mathrm{d}}^{n}+t_{\mathrm{d}}^{n+1}\right)/2\right]$
is different form one.

\textbf{Parameters used: }\emph{Figure \ref{fig:traces_and_dist}:}
$N=1000,\ \rin\tau=0.14,\ \rout\tau=0.07,\ \tau=\sqrt{\tm\ts}=14\mathrm{msec,\ }\tm/\ts=8,\ T/\tau N=1.4\,$.
\emph{Figure \ref{fig:jitter_and_eps}:} In (b) and (c) $N=1000,\ \rin=10\mathrm{Hz},$
$\rout=5\,\mathrm{Hz},\ \tau=10\,\mathrm{msec},\ \tm=20$~msec, $T=12\,\mathrm{sec}$.\textbf{
}\emph{Figure \ref{fig:sv_dist}:} In all panels $\tau=14\,\mathrm{msec},\ \tm=40\,\mathrm{msec},\ \varepsilon=\mbox{\ensuremath{\tau}},\ T/N\tau=2,\ N=1000$.
In (a)-(b) $\rin=10\mathrm{Hz}.$ In (c)-(d) $\rin=20\mathrm{Hz}$,
`target' pattern consist of Poisson input spikes of duration $2\tau$.
`null' pattern is the `target' patern were 10\% of the spikes were
randomly modified. \emph{Figure \ref{fig:Temporal-XOR}:}\textbf{
}$\tm=10\mbox{msec}$, $\ts=5\mathrm{msec}$, $\varepsilon=5\mathrm{msec}$,
$\Delta=5\mathrm{msec}$. \emph{Figure \ref{fig:Neuronal-architecture-for-nonlinear-kernels}:}\textbf{
}$N=50$, $T=0.5\mathrm{sec}$, $\rin=20\mathrm{Hz}$, $\tm=20\mbox{msec}$,
$\ts=5\mathrm{msec}$, $\varepsilon=10\mathrm{msec}$.

\global\long\def\tm{\tau_{\mathrm{m}}}%
 
\global\long\def\ts{\tau_{\mathrm{s}}}%
 
\global\long\def\rout{r_{\mathrm{out}}}%
 
\global\long\def\rin{r_{\mathrm{in}}}%
 
\global\long\def\td{t_{\mathrm{d}}}%

\title{Supplementary Methods: Temporal Support Vectors for Spiking Neural
Networks}
\author{Ran Rubin and Haim Sompolinsky}
\date{}
\maketitle
\tableofcontents{}

\pagebreak{}

\section{Existence of Positive Dynamic Margin}

In this section we will show that for any $\boldsymbol{\omega}$ which
satisfies the requirements of the task the dynamic margin is always
strictly positive given that $\boldsymbol{x}\left(t\right)$ and $\mu\left(t\right)$
are well behaved (see below).

\textbf{Theorem:} Given the inputs $\boldsymbol{x}\left(t\right)$,
desired times $\left\{ \td\right\} $, profile of the dynamic margin
$\mu\left(t\right)$, weight vector\textbf{ $\boldsymbol{\omega}$
}and the following:
\begin{enumerate}
\item $\boldsymbol{\omega}$ is a solution, \emph{i.e. }for all $t_{\mathrm{d}}\in\left\{ \td\right\} $\emph{,
$U\left(\td\right)=\theta,$ $\frac{\mathrm{d}U}{\mathrm{d}t}\left(\td\right)>0$
}and $U\left(t\right)<\theta$ for all $t\ne\td$.
\item $U\left(t\right)$ and $\mu\left(t\right)$ are continuous between
$\td$'s
\item $\mu\left(t\right)>0$ for $t\ne\td$, $\mu\left(\td\right)=0$, $\frac{\mathrm{d}\mu}{\mathrm{d}t}\left(\td\right)>-\infty$,
$\max_{t}\mu\left(t\right)=1$.
\end{enumerate}
Then the dynamic margin, $\delta$, is greater then zero.

\textbf{Proof:} From the definition of the dynamic margin of a given
solution, $\boldsymbol{\omega}$ (Eq. 2 in the main text), we have
that $\theta-U\left(t\right)\ge\left\Vert \boldsymbol{\omega}\right\Vert \delta\mu\left(t\right)$,
where $\delta$ is the maximal constant for which this inequality
is satisfied for all $t\ne\td$. To show that $\delta>0$ we will
bound it from below using a constant $\epsilon>0$ that satisfies
\begin{equation}
\theta-U\left(t\right)\ge\epsilon\mu\left(t\right)
\end{equation}
for all $t$. This ensures that $\delta\ge\frac{\epsilon}{\left\Vert \boldsymbol{\omega}\right\Vert }>0$.

For every $\td$ we define, 
\begin{equation}
\epsilon_{\td}\equiv\min_{t}\frac{\theta-U\left(t\right)}{\td-t}/\max_{t}\frac{\mu\left(t\right)}{\td-t}
\end{equation}
where the minimum and maximum are taken over $t\in\left(\td-\Delta_{\td},\td\right)$
with some $\Delta_{\td}>0$ smaller then the inter-spike-interval
between $\td$ and the desired time before it. Both the nominator
and the denominator are positive and the denominator is finite. Thus
$\epsilon_{\td}>0$. 

In addition we have 
\begin{align}
\frac{\theta-U\left(t\right)}{\td-t} & \ge\min_{t^{\prime}}\frac{\theta-U\left(t^{\prime}\right)}{\td-t^{\prime}}\\
\frac{\mu\left(t\right)}{\td-t} & \le\max_{t^{\prime}}\frac{\mu\left(t^{\prime}\right)}{\td-t^{\prime}}\label{eq:4}
\end{align}
Multiplying both sides of (\ref{eq:4}) by $\epsilon_{\td}$ we find
that for all $t\in\left(\td-\Delta_{\td},\td\right)$, 
\begin{equation}
\theta-U\left(t\right)\ge\epsilon_{\td}\mu\left(t\right)\ .
\end{equation}

We define 
\begin{equation}
\epsilon_{1}=\min_{\td}\epsilon_{\td}\ .
\end{equation}

Now, for every $t\notin\bigcup_{\td}\left(\td-\Delta_{\td},\td\right]$
we have $U\left(t\right)<\theta$ therefore there exists $\epsilon_{2}>0$
such that for these times: 
\begin{equation}
\theta-U\left(t\right)\ge\epsilon_{2}\ge\epsilon_{2}\mu\left(t\right)
\end{equation}
since $\mu$$\left(t\right)>0$ and the maximal value of $\mu\left(t\right)$
is set to $1$. 

Finally we define:
\begin{equation}
\epsilon=\min\left(\epsilon_{1},\epsilon_{2}\right)>0\ ,
\end{equation}
 and we have that for all times $t$ 
\begin{equation}
\theta-U\left(t\right)\ge\epsilon\mu\left(t\right)\ _{\blacksquare}
\end{equation}

\section{Maximizing the Dynamic Margin}

To maximize the dynamic margin we minimize $\frac{1}{2}\boldsymbol{\omega}^{T}\boldsymbol{\omega}$
under the constraints (c.I) $-\left(U\left(t\right)-\theta\right)\ge\mu\left(t\right)$
for all $t\ne t_{\mathrm{d}}$, (c.II) $U\left(t_{\mathrm{d}}\right)=\theta$
and (c.III) $\frac{\mathrm{d}U}{\mathrm{d}t}\left(t_{\mathrm{d}}\right)\ge-\frac{\mathrm{d}\mu}{\mathrm{d}t}\left(t_{\mathrm{d}}\right)$.
We define the Lagrangian:
\begin{equation}
\mathcal{L}=\frac{1}{2}\boldsymbol{\omega}^{T}\boldsymbol{\omega}-\int_{0}^{T}\mathrm{dt}\alpha\left(t\right)\left[-\left(U\left(t\right)-\theta\right)-\mu\left(t\right)\right]-\sum_{\td}\gamma_{\td}\left[\frac{\mathrm{d}U}{\mathrm{d}t}\left(t_{\mathrm{d}}\right)+\frac{\mathrm{d}\mu}{\mathrm{d}t}\left(t_{\mathrm{d}}\right)\right]-\sum_{\td}\beta_{\td}\left[U\left(\td\right)-\theta\right]\ ,\label{eq:lagrangian}
\end{equation}
and our task is to minimize $\mathcal{L}$ w.r.t. $\boldsymbol{\omega}$
and $\theta$, and maximize $\mathcal{L}$ w.r.t. $\alpha\left(t\right)$,
$\beta_{\td}$ and $\gamma_{\td}$ under the constraints: $\alpha\left(t\right)\ge0$
and $\gamma_{\td}\ge0$. The Karush-Kuhn-Tucker theorem states that,
for the optimal values of the coefficients, constraints (c.I)-c.(III)
are satisfied and the following hold:
\begin{eqnarray}
\forall t\ne\td\ \alpha\left(t\right)\left[-\left(U\left(t\right)-\theta\right)-\mu\left(t\right)\right] & = & 0\label{eq:KKT}\\
\alpha\left(t\right) & \ge & 0\nonumber \\
\forall\td\ \gamma_{\td}\left[\frac{\mathrm{d}U}{\mathrm{d}t}\left(t_{\mathrm{d}}\right)+\frac{\mathrm{d}\mu}{\mathrm{d}t}\left(t_{\mathrm{d}}\right)\right] & = & 0\nonumber \\
\gamma_{\td} & \ge & 0\nonumber 
\end{eqnarray}
That is $\alpha\left(t\right)$ and $\gamma_{\td}$ can be non zero
only at times in which the constraints (c.I) and (c.III) respectively
are satisfied with an equality. We assume that for a generic $U\left(t\right)$
and $\mu\left(t\right)$ the analytical shape of $U\left(t\right)+\mu\left(t\right)$
does not allow for a constant value for any non infinitesimal duration.
Thus, it follows that $\alpha\left(t\right)$ can be non-zero only
at isolated local maxima of $U\left(t\right)+\mu\left(t\right)$ in
which $U\left(t\right)=\theta-\mu\left(t\right)$. $\alpha\left(t\right)$
can be written as 
\begin{equation}
\alpha\left(t\right)=\sum_{t_{\mathrm{SV}}}\alpha_{t_{\mathrm{SV}}}\delta\left(t-t_{\mathrm{SV}}\right)\label{eq:discrete_alpha}
\end{equation}
where $\delta\left(t\right)$ is the Dirac delta function.

Taking the derivative of $\mathcal{L}$ w.r.t. $\boldsymbol{\omega}^{T}$
and equating to zero, together with (\ref{eq:discrete_alpha}) leads
to the form of the optimal weight vector $\boldsymbol{\omega}^{\star}$
given by eq. 3 in the main text. 

It is important to note that despite the fact that constraints (c.I)
and (c.II) implicitly enforce constraint (c.III), we find that explicitly
including (c.III) in $\mathcal{L}$ is required for the application
of the Karush-Kuhn-Tucker theorem for this case. In fact, without
including the slope constraints in $\mathcal{L}$, it is possible
to construct simple examples in which the solution cannot be expressed
with $\alpha\left(t\right)$ in the form of (\ref{eq:discrete_alpha}). 

\section{Effective Perceptron Theory for the Dynamic Margin and Support Vectors}

The capacity of the LIF neuron to realize random input-output transformations
and its dependance on the various parameters was studied extensively,
numerically and analytically, in \cite{memmesheimer_learning_2014}.
It was found that the relevant time constant of the neuronal dynamics
is $\tau=\sqrt{\tm\ts}$ and that the duration of the longest implementable
input-output transformation, $T_{c}$, is extensive and that $T_{c}/N\tau$
only depends on the mean number of desired output spike within time
$\tau$, $\rout\tau$. In addition, for $\rout\tau\lesssim0.1$, the
system is well approximated by an effective Perceptron (see below).
This allows for the evaluation of the capacity and of statistical
properties of the solutions below the capacity, using the Gardner
theory of the Perceptron \cite{gardner_space_1988}. Further, the
above analogy with the Perceptron can also be used to investigate
the change in $\delta^{\star}$ and the properties of the SV's as
the load, i.e., the duration of the spike sequence, $T$, increases.
The predicted mean $\delta^{\star}$ and mean number of SV's, as well
as the numerical simulations are shown in Figure 5(a)-(b). 

\emph{Effective Perceptron Approximation (Gardner Theory):} We approximate
the spiking neuron as an effective Perceptron with $N_{\mathrm{eff}}=N-\rout T$
input synapses, classifying $P_{\mathrm{eff}}=T\left(1/\tau+\rout\right)$
uncorrelated patterns, $\rout T$ patterns labeled as `target` patterns
and $T/\tau$ patterns labeled as `null` patterns. The capacity, mean
margin, and mean number of support vectors, $n_{\mathrm{SV}}$, are
a function of the effective load $\alpha_{\mathrm{eff}}=\frac{P_{\mathrm{eff}}}{N_{\mathrm{eff}}}$,
and the ratio between the effective number of `target` patterns and
$P_{\mathrm{eff}}$, $f_{\mathrm{eff}}=\rout\tau/\left(1+\rout\tau\right)$.
In practice we derive expressions for $\alpha_{\mathrm{eff}}$ and
$n_{\mathrm{SV}}$ as a function of the margin in units of the standard
deviation of the inputs, $\kappa=\frac{\delta^{\star}}{\mathrm{std}\left(x_{i}\left(t\right)\right)}$
and $f_{\mathrm{eff}}$:
\begin{align}
\alpha_{\mathrm{eff}}\left(\kappa,f_{\mathrm{eff}}\right) & =\left[f_{\mathrm{eff}}\int_{\eta-\kappa}^{\infty}Dt\left(t-\eta+\kappa\right)^{2}+\left(1-f_{\mathrm{eff}}\right)\int_{-\eta-\kappa}^{\infty}Dt\left(t+\eta+\kappa\right)^{2}\right]^{-1}\\
n_{\mathrm{SV}}\left(\kappa,f_{\mathrm{eff}}\right) & =P_{\mathrm{eff}}\left[f_{\mathrm{eff}}H\left(\eta-\kappa\right)+\left(1-f_{\mathrm{eff}}\right)H\left(-\eta-\kappa\right)\right]
\end{align}
where $Dt=\frac{e^{-\frac{t^{2}}{2}}}{\sqrt{2\pi}}\mathrm{d}t$, $H\left(y\right)=\int_{y}^{\infty}Dt$,
and $\eta$ is given by the solution to
\begin{equation}
f_{\mathrm{eff}}\int_{\eta-\kappa}^{\infty}Dt\left(t-\eta+\kappa\right)=\left(1-f_{\mathrm{eff}}\right)\int_{-\eta-\kappa}^{\infty}Dt\left(t+\eta+\kappa\right)\ .
\end{equation}
Simple algebra can be used to then find the relation between $\delta^{\star}$,
$\mbox{\ensuremath{\tau}}n_{\mathrm{SV}}/T$ and $T/\tau N$.

\section{Dual Lagrangian and Application of Kernel Method for T-SVM}

To apply the Kernel method to T-SVM we enforce the slope constraints
(c.III) implicitly by excluding from the temporal integration in (\ref{eq:lagrangian})
segments of duration $\Delta$ prior to every $\td$ and removing
the Lagrange coefficients enforcing the slope constraints, $\left\{ \gamma_{\td}\right\} $.
In this case, the K.K.T theorem implies that the optimal value of
$\alpha\left(t\right)$ satisfies equations (\ref{eq:KKT}) and (\ref{eq:discrete_alpha})
where $t\in\left\{ t_{\mathrm{SV}}\right\} $ only if $t$ is an isolated
local maxima of $U\left(t\right)+\mu\left(t\right)$ in which $U\left(t\right)=\theta-\mu\left(t\right)$
\textbf{or} if $t=\td-\Delta$. In the $\Delta\rightarrow0$ limit
this exclusion is equivalent to explicitly enforcing constraints (c.III).
In practice our simulation of nonlinear kernels are approximated with
discrete time steps of size $\Delta t$, and we use $\Delta=\Delta t$.

To apply the kernel method we derive the dual Lagrangian. Since our
algorithm uses a sequence of sets of sampled times we consider the
primal Lagrangian for a given set $\left\{ t_{s}\right\} $ and a
set of desired times $\text{\ensuremath{\left\{  \td\right\} } }$:
\begin{equation}
\mathcal{L}=\frac{1}{2}\boldsymbol{\omega}^{T}\boldsymbol{\omega}-\sum_{t_{s}}\alpha_{t_{s}}\left[-\left(U\left(t_{s}\right)-\theta\right)-\mu\left(t_{s}\right)\right]-\sum_{\td}\beta_{\td}\left[U\left(\td\right)-\theta\right]\ .
\end{equation}

The dual Lagrangian is obtained by minimizing $\mathcal{L}$ w.r.t.
$\boldsymbol{\omega}^{T}$ and expressing the Lagrangian using the
optimal weight vector:
\begin{align}
\mathcal{Q}= & -\frac{1}{2}\sum_{t_{s},t_{s}^{\prime}}\alpha_{t_{s}}K\left[\boldsymbol{x}\left(t_{s}\right),\boldsymbol{x}\left(t_{s}^{\prime}\right)\right]\alpha_{t_{s}^{\prime}}-\frac{1}{2}\sum_{\td,\td^{\prime}}\beta_{\td}K\left[\boldsymbol{x}\left(\td\right),\boldsymbol{x}\left(\td^{\prime}\right)\right]\beta_{\td^{\prime}}\\
 & -\sum_{t_{s},\td}\alpha_{t_{s}}K\left[\boldsymbol{x}\left(t_{s}\right),\boldsymbol{x}\left(\td\right)\right]\beta_{\td}+\sum_{t_{s}}\alpha_{t_{s}}\mu\left(t_{s}\right)\nonumber \\
 & +\theta\left[\beta_{\td}\left(1+x_{\mathrm{reset}}\left(\td\right)\right)-\sum_{t_{s}}\alpha_{t_{s}}\left(1+x_{\mathbb{\mathrm{reset}}}\left(t_{s}\right)\right)\right]\nonumber 
\end{align}
where $K\left[\boldsymbol{u},\boldsymbol{v}\right]=\boldsymbol{u}^{T}\boldsymbol{v}$.
To solve the optimization problem we maximize $\mathcal{Q}$ w.r.t
$\alpha_{t_{s}}$ and $\beta_{\td}$ under the constraints:
\begin{align}
\alpha_{t_{s}} & \ge0\\
\beta_{\td}\left(1+x_{\mathrm{reset}}\left(\td\right)\right)-\sum_{t_{s}}\alpha_{t_{s}}\left(1+x_{\mathbb{\mathrm{reset}}}\left(t_{s}\right)\right) & =0\nonumber 
\end{align}
In practice, $\beta_{\td}$ and $\theta$ can be found analytically
and SVM optimization can be performed only w.r.t the $\alpha_{t_{s}}$
coefficients.

For nonlinear kernels one simply replaces the linear kernel $K\left[\boldsymbol{u},\boldsymbol{v}\right]$
in $\mathcal{Q}$, with a nonlinear, symmetric, positive definite
kernel $\mathcal{K}\left[\boldsymbol{u},\boldsymbol{v}\right]$. 

\section{Extensions}

\subsection{Dynamic Margin for Non-Spiking Dynamical Systems}

The application of T-SVM is not limited to spiking neurons only. Dynamic
margin maximization can be applied to any dynamical system in which
decision is based on threshold crossing. 

We consider a binary dynamical unit, whose output, $s\left(t\right)$,
is given by $s\left(t\right)=\mathrm{sign}\left(U\left(t\right)-\theta\right),$
where $\theta$ and $U\left(t\right)$ are the units threshold and
internal dynamical variable analog to the membrane potential in the
spiking neurons, respectively. The unit linearly sums $N$ input afferents
defined by $N$ time varying signals, $x_{i}\left(t\right),\ i=1,2,\dots,N$.
For a set of weights, $\omega_{i},\ i=1,2,\dots,N$, $U\left(t\right)$
is defined by 
\begin{eqnarray}
U\left(t\right) & = & \boldsymbol{\omega}^{T}\boldsymbol{x}\left(t\right)\ ,\label{eq:membrane_potential}
\end{eqnarray}
where $\boldsymbol{\omega}$ and $\boldsymbol{x}\left(t\right)$ are
the $N$-dimensional vectors of the weights and inputs at time $t$,
respectively.

The training data consists of an input trace vector, $\boldsymbol{x}\left(t\right)$,
and a desired output function, $y\left(t\right)=\pm1$ which is assumed
to be piece-wise constant, for all $t\in\left[0,T\right]$. $y\left(t\right)$
defines a discrete set of desired threshold crossing times, $\left\{ t_{\mathrm{d}}\right\} $,
at which $y\left(t\right)$ changes sign. The task of learning is
to modify $\boldsymbol{\omega}$ such that the unit's output equals
the desired output, \emph{i.e.}, $s\left(t\right)=y\left(t\right)$
for all $t\ne t_{\mathrm{d}}$. Thus, $U\left(t\right)$ is required
to be above $\theta$ when $y\left(t\right)=1$, below $\theta$ when
$y\left(t\right)=-1$ and cross $\theta$ precisely at the desired
times, $\left\{ t_{\mathrm{d}}\right\} $.

As in the case of spiking neurons, simple margin maximization cannot
be applied to these kind of systems due to the continuous threshold
crossing at $\td$. However it is possible to generalize the definition
of the dynamic margin for this case. The dynamic margin of a weight
vector is defined as: 
\begin{equation}
\delta=\min_{t\ne\td}\frac{y\left(t\right)\left(U(t)-\theta\right)}{\left\Vert \boldsymbol{\omega}\right\Vert \mu\left(t\right)}\ ,\label{eq:dynamic_margin_def}
\end{equation}
where here we require that $\mu(t)$ is a continuous function that
satisfies: $\mu\left(t\right)>0$ for $t\ne\td$, $\mu\left(\td\right)=0$,
$\frac{\mathrm{d}\mu}{\mathrm{d}t}\left(\td^{-}\right)>-\infty$,
$\frac{\mathrm{d}\mu}{\mathrm{d}t}\left(\td^{+}\right)<\infty$ and
$\max_{t}\mu\left(t\right)=1$. As before, we define the optimal solution,
$\boldsymbol{\omega}^{\star}$, as the weight vector that possesses
the maximal dynamic margin, $\delta^{\star}$.

Maximizing the dynamic margin in this case is equivalent to minimizing
the norm of $\boldsymbol{\omega}$ under the constraints (c.I) $y\left(t\right)\left(U\left(t\right)-\theta\right)\ge\mu\left(t\right)$
for all $t\ne t_{\mathrm{d}}$; (c.II) $U\left(t_{\mathrm{d}}\right)=\theta$;
(c.III) $y\left(t_{\mathrm{d}}^{-}\right)\frac{\mathrm{d}U}{\mathrm{d}t}\left(t_{\mathrm{d}}^{-}\right)\le\frac{\mathrm{d}\mu}{\mathrm{d}t}\left(t_{\mathrm{d}}^{-}\right)$
and (c.IV) $y\left(t_{\mathrm{d}}^{+}\right)\frac{\mathrm{d}U}{\mathrm{d}t}\left(t_{\mathrm{d}}^{+}\right)\ge\frac{\mathrm{d}\mu}{\mathrm{d}t}\left(t_{\mathrm{d}}^{+}\right)$.
Note that for continuous inputs, imposing (c.I) fulfills all the requirements
of the task and implicitly enforces the remaining constraints. However
as in the spiking neurons case, for the application of the K.K.T theorem
constraints (c.II)-(c.IV) must be explicitly included in the optimization
as explained above. T-SVM algorithm, and the Kernel method can be
applied to these type of systems in a similar manner as to spiking
neurons. 

\subsection{Learning Optimal Temporal Filters}

An example of system in which optimal temporal filters can be learned
such is 
\begin{equation}
U\left(t\right)=\int_{0}^{\tau_{\max}}\mathrm{d}\tau\boldsymbol{\omega}\left(\tau\right)^{T}\boldsymbol{x}\left(t-\tau\right)\ .
\end{equation}
where $\boldsymbol{\omega_{i}}\left(\tau\right)$ is the synaptic
temporal filter of the $i-$th afferent. In this case we define the
dynamic margin according to (\ref{eq:dynamic_margin_def}) with $\left\Vert \boldsymbol{\omega}\right\Vert =\left[\int_{0}^{\tau_{\max}}\mathrm{d}\tau\boldsymbol{\omega}\left(\tau\right)^{T}\boldsymbol{\omega}\left(\tau\right)\right]^{\frac{1}{2}}$,
and maximizing it can be done in the same manner as before. The optimal
solution $\boldsymbol{\omega}\left(\tau\right)$ is a sum of a set
of select input\textit{ traces}: $\boldsymbol{x}\left(t_{\mathrm{d}}-\tau\right)$,
$\frac{\mathrm{d}\boldsymbol{x}}{\mathrm{d}t}\left(\td-\tau\right)$
, and $\boldsymbol{x}\left(t_{\mathrm{SV}}-\tau\right)$ and with
coefficients $\beta_{t_{\mathrm{d}}}$, $-\gamma_{\td}y\left(t_{\mathrm{\mathring{d}}}^{-}\right),$
and $\alpha_{t_{\mathrm{SV}}}y\left(t_{\mathrm{SV}}\right)$, respectively.

\end{document}